\documentclass[pre,article]{revtex4}
\usepackage{graphicx}
\usepackage{amsmath}

\providecommand{\U}[1]{\protect\rule{.1in}{.1in}}

\begin{document}

\title{\ Is the Tsallis $q$-mean value instable? }
\author{A. Cabo}
\affiliation{\textit{Theoretical Physics Department, Instituto de Cibern\'{e}tica,
Matem\'{a}tica y F\'{i}sica, Calle E, No. 309, Vedado, La Habana, Cuba.}}
\begin{abstract}
\noindent  \hspace{0in} \hspace{0in}The recent argue about the \hspace{0in}%
existence of an instability in the definition of the \hspace{0in}mean value
appearing in the Tsallis non extensive Statistical Mechanic is reconsidered.
\hspace{0in} Here, it is simply underlined that the pair of probability
distributions employed in constructing the instability statement \hspace{0in}%
have  a discontinuous \hspace{0in}limit when the number of states tends to infinity.
\hspace{0in}That is, although for an arbitrary but finite  number of states W  both
probability distributions are normalized to the unit, their  limits $W->\infty $ do
not satisfy the normalization condition and thus are not allowed $escort$
probabilities for the q-mean value. However, \hspace{0in}  similar  distributions
converging to the former ones when a parameter $W_o$ is tending to infinity are
defined here. They both  satisfy the normalization to the unity in the limit
$W\rightarrow \infty $. This simple change allows to  show that the stability
condition becomes satisfied, for whatever large but fixed value of $W_o$ is chosen.

\bigskip
\end{abstract}

\maketitle

\smallskip

Nowadays, the investigation about the description of the non equilibrium statistical
properties of a large number of physical systems is being considered by employing
\hspace{0in}the methods of non extensive Statistical Physics introduced by C. Tsallis
\cite{tsallis}. The successes of this theory  had been many. The
\hspace{0in}description has been able to predict the  deviations from the
\hspace{0in}Boltzman distributions \hspace{0in} measured for \hspace{0in}numerous
physical systems and motivated an intense research  activity seeking of new
statistical approaches for non equilibrium systems \cite
{tsallis,driscol,boghosian,smith,curilef, carlos}.

\smallskip \hspace{0in}Recently,  however, there had been some argues in the
literature(\cite{abe1,abe2,abe3,lutsko,hanel}) supporting the existence of an \hspace{0in}%
instability in a basic quantity  defining the \hspace{0in}theory: the special Tsallis
mean \hspace{0in}value ($q$-mean value) of an arbitrary physical quantity $Q$ as
given by
\begin{eqnarray}
\langle Q\rangle _q &=&\frac{\sum_{i=1}^Wp_i^qQ_i}{\sum_{i=1}^Wp_i^q}, \\
\sum_{i=1}^Wp_i &=&1,
\end{eqnarray}
which leads to the usual one in the $q\rightarrow 1$ limit $\langle Q\rangle
_1=\sum^W_ip_iQ_i.$ \hspace{0in}In the above formula, $W$ is the number of states in
the system when it is a finite number and the $p_i$ are the so  called $escort$
probabilities.

Basically,  those works  claim  that there exist particular probability
perturbations, which are properly defined for finite systems ($W$ being
finite), under   which the $q-$mean values of the physical quantities in the
infinite system limit, $W\rightarrow \infty ,$ show variations that can not
be made arbitrarily small, when the perturbations (also taken in the limit
of infinite system) of the probabilities are chosen as sufficiently small
\cite{abe1}. \hspace{0in} \hspace{0in}

In this note we will reconsider the discussion given in Ref. \cite{abe1}.
\hspace{0in} We will follow the notation of that work and define the two
probability distributions $\{p_i\}$and $\{p_i^{\prime }\}$ for $i=1,2,....W,$
for a system having \hspace{0in}$W$ states. The perturbation of the
probability \hspace{0in}will be defined by the sequence of differences
\begin{equation}
\Delta p_i=p_i-p_i^{\prime },\text{ \hspace{0in} \hspace{0in} \hspace{0in}
\hspace{0in}}i=1,2,....W.
\end{equation}

\smallskip

\hspace{0in}For starting, it is helpful to  state, that the demonstration given in
\cite{abe1} of the ''non small'' character of the variation of the $q$-mean value of
a given physical quantity $Q$ (under sufficiently small perturbations of the
probabilities in the infinite number of states limit) is \hspace{0in}mathematically
correct. However, in spite of this, it should be stressed that the instability result
is claimed to be valid for the infinite system $\hspace{0in}(W\rightarrow \infty ).$
Therefore, \hspace{0in}it must be pointed out that  a necessary requirement
\hspace{0in}for the   instability to exist is that in the infinite number
states limit $W\rightarrow \infty ,$ \hspace{0in}%
both distributions should retain  their essential meaning of being proper
$escort$ probabilities \hspace{0in} satisfying $\sum_{i=1}^\infty p_i=1$ and $%
\sum_{i=1}^\infty p_i^{\prime }=1.$ \hspace{0in}This becomes necessary \hspace{0in%
}because the infinite number of state system is well defined, and the limit
$W\rightarrow \infty$  of also well defined $escort$ probabilities should satisfy the
normalization condition, in order to be considered as acceptable $escort$
probabilities for the infinite systems. \hspace{0in}The lack of satisfaction of this
condition in the instability argue in Ref. \cite{abe1}
 is the essential point claimed in the present letter.

In what follows, it will be noticed that this central requirement is drastically
violated by the probability distributions \hspace{0in}considered in Ref. \cite{abe1}.
Further,  it will shown that similarly defined distributions, but slightly modified
for satisfying the above posed additional condition, \hspace{0in}directly imply the
satisfaction of the stability condition for the infinite system.

The fact that the infinite $W$ \hspace{0in} limit \hspace{0in}destroys the well
defined character of the distributions defined in Ref. \cite{abe1}, follows form a
simple inspection of the formulae for those quantities in the $W\rightarrow \infty $
limit. For the case  $0<q<1,$ the limits are
\begin{equation}
p_i=\delta _{i1},\text{ \hspace{0in} \hspace{0in} }p_i^{\prime }=(1-\frac
\delta 2)\text{ }p_i,\text{ \hspace{0in} }i=1,2,....,\infty ,
\end{equation}
and for the case $q>1:$

\hspace{0in} \hspace{0in}
\begin{equation}
p_i=0,\text{ \hspace{0in} \hspace{0in} }p_i^{\prime }=\frac {\delta} {2}
\delta_{i1},\text{ \hspace{0in} }i=1,2,....,\infty .
\end{equation}

\smallskip

\smallskip Its is clear from the above expressions that for any value of the
parameter $\hspace{0in}$ $\delta ,$ only one of the four distributions
satisfies the normalization condition for the $escort$ probabilities of the
infinite system. Therefore, we can conclude that non of the two sets of
probability distributions obtained as the limits of the (well defined)
finite systems distributions, can be associated to \hspace{0in}proper
perturbations of the Tsallis mean values for the infinite physical system.

\smallskip Let us now show that the when the \hspace{0in}distributions for
finite systems employed in Ref. \cite{abe1} are slightly modified to satisfy the
condition of leading to allowed $escort$ probabilities, then the \hspace{0in}
stability condition is satisfied for the \hspace{0in}redefined probability
perturbations in  the infinite system. This will complete the issues aimed to be
discussed in this letter.

For this purpose, let us consider, the variations of the $q-$mean value of a physical
quantity. The construction will  closely follows the one in Ref.  \cite{abe1}.
\hspace{0in}Consider a large finite value of $W$ \ and another also large number
$W_o<W$. \ \ The number of states of the system $W$ will be taken in the infinite
limit, but $W_o,$ although
having an arbitrarily large value will  remain to be fixed when $%
W\rightarrow \infty $. \ \ The new pairs of probability distribution to be defined
will essentially coincide in their analytic expressions with the ones given in Ref.
\cite{abe1} but in which $W$ will be replaced by $W_o$. \hspace{0in} \ \ The concrete
expressions for the case  $0<q<1$ are
\begin{eqnarray}
p_i &=&\delta _{i1},\text{ \hspace{0in} \hspace{0in} }p_i^{\prime }=(1-\frac
\delta 2\frac{W_o}{W_o-1})p_i+\frac \delta 2\frac{\theta (W_o-i)}{W_o-1},%
\text{ \hspace{0in} }i=1,2,....,W, \\
\theta (n) &=&\left\{
\begin{tabular}{l}
$1,\ \ \ \ \ n\geq 0$ \\
$0,\ \ \ \ \ n<0$%
\end{tabular}
\right. ,n-{integer},
\end{eqnarray}
and for $q>1$:
\begin{equation}
p_i=\frac{(1-\delta _{i1})\theta (W_o-i)}{W_o-1},\text{ \hspace{0in} \hspace{%
0in} }p_i^{\prime }=(1-\frac \delta 2)p_i+\frac \delta 2\delta _{i1},\text{
\hspace{0in} }i=1,2,....,W.
\end{equation}

\ It can be noted that these definitions by construction, do not depend on the number
of the states of the system $W.$ \hspace{0in} \hspace{0in} Therefore, all the four
distributions  satisfy the normalization conditions
\begin{equation}
\sum_{i=1}^Wp_i=1,\text{ \ \ \ }\sum_{i=1}^Wp_i^{\prime }=1,
\end{equation}
for arbitrary values of $W_o$ and $W$. \ Moreover,  these distributions, due
to their independence of $W,$ exactly coincide with their limit for $%
W\rightarrow \infty $ . Thus, they remain exactly normalized to the unity
for the infinite number of states limit, and thus  correspond to valid $%
escort$ probabilities for the $q-$mean value when the \ system already has
an infinite number of states. \ \ Exactly in the same way as it is derived
in\ Ref. [\cite{abe1}], the modulus of the difference between the $q$- mean
values evaluated with the probability sequences $\{p_i\}$ and $\{p_i^{\prime
}\},$for the range $0<q<1$, can be \ directly calculated to be
\begin{equation}
\mid \langle Q\rangle _q-\langle Q\rangle _q^{\prime }\mid ={\huge \mid }Q_1-%
\frac{(1-\frac \delta 2)^qQ_1+\frac{(\frac \delta 2)^q}{(W_o-1)^q}%
\sum_{i=2}^{W_o}Q_i}{(1-\frac \delta 2)^q+(\frac \delta 2)^q(W_o-1)^{1-q}}%
{\huge \mid .}
\end{equation}

This expression is $W$ independent as noted before, then its $W\rightarrow
\infty $ limit coincides with it. Now, it is clear that the above formula
implies that $\mid \langle Q\rangle _q-\langle Q\rangle _q^{\prime }\mid $
can be fixed to be smaller \ than any arbitrarily small quantity $\epsilon $
\ after also selecting $\delta $ to be sufficiently small for any
arbitrarily large but fixed constant $W_o$. \ Therefore, the slight
redefinition of the densities for to be well defined $escort$ probabilities
for $W\rightarrow \infty $ had allowed to satisfy the stability condition.

Finally, in a similar way, the perturbation of the mean values for the $q>1$ \ case,
can be expressed  in identical form as in Ref. \cite{abe1} but by replacing $W$
\hspace{0in} by $W_o$ , \ to \ get the expression
\begin{eqnarray}
\mid \langle Q\rangle _q-\langle Q\rangle _q^{\prime }\mid &=&{\huge \mid }%
\frac{\overline{Q}-Q_1}{W_o-1}-\frac{\frac{(1-\frac \delta 2)^q}{(W_o-1)^q}%
(W_o\overline{Q}-Q_1)+(\frac \delta 2)^qQ_1}{(\frac \delta 2)^q+(1-\frac \delta
2)^q(W_o-1)^{1-q}}{\huge \mid }, \\
\overline{Q} &=&  \frac{\sum_{i=1}^{W_o}Q_i}{W_o}.
\end{eqnarray}

Again, this formula coincides with its $W\rightarrow \infty $ limit and,  in
addition,  it also can be reduced to be smaller than any arbitrarily small
quantity $\epsilon $ by\ choosing the parameter $\delta $ as appropriately
small in size.

The above remarks, \ establish the stability conditions for the $q-$mean value for a
class of \ disturbances which are closely similar to the ones employed in \ Ref.
\cite{abe1} , but slightly changed  in order to satisfy
the normalization conditions for the $escort$ probabilities in the $\hspace{%
0in}W\rightarrow \infty $ limit.

\smallskip The author acknowledge the support received from the Proyecto
Nacional de Ciencias B\'asicas (PNCB, CITMA, Cuba) and from the Network N-35 of the
Office of External Activities (OEA) of the ICTP (Italy).

\end{document}